# A hydroxamic acid-methacrylated collagen conjugate for the modulation of inflammation-related MMP upregulation

He Liang,[1, 2] Stephen J. Russell,[1] David, J. Wood,[2] Giuseppe Tronci[1,2*]

[1] Clothworkers' Centre for Textile Materials Innovation for Healthcare, School of Design, University of Leeds, United Kingdom

[2] Biomaterials and Tissue Engineering Research Group, School of Dentistry, St. James's University Hospital, University of Leeds, United Kingdom

* Email correspondence: g.tronci@leeds.ac.uk

## Abstract

Medical devices with matrix metalloproteinase (MMP)-modulating functionality are highly desirable to restore tissue homeostasis in critical inflammation states, such as chronic wounds, rotator cuff tears and cancer. The introduction of MMP-modulating functionality in such devices is typically achieved via loading of either rapidly-diffusing chelating factors, e.g. EDTA, or MMP-cleavable substrates, raising issues in terms of non-controllable pharmacokinetics and enzymatic degradability, respectively. Aiming to accomplish inherent, long-term, device-induced MMP regulation, this study investigated the synthesis of a hydroxamic acid (HA)-methacrylated collagen conjugate as the building block of a soluble factor-free MMP-modulating hydrogel network with controlled enzymatic degradability. This was realised via a two-step synthetic route: (i) type I collagen was functionalised with photonetwork-inducing methacrylic anhydride (MA) adducts in the presence of triethylamine (TEA); (ii) this methacrylated product was activated with a water-soluble carbodiimide prior to reaction with hydroxylamine, resulting in MMP-chelating HA functions. Nearly-quantitative methacrylation of collagen amines was observed via 2,4,6-trinitrobenzenesulfonic acid (TNBS) assay; this was key to avoiding intramolecular crosslinking side reactions during carbodiimide-mediated activation of collagen carboxyl groups. The molar content of HA adducts was indirectly quantified via conversion of remaining carboxyl functions into ethylenediamine (EDA), so that 12-16 mol.% HA was revealed in the conjugate by both TNBS and Ninhydrin assays. Resulting UV-cured, HA-bearing collagen hydrogels proved to induce up to ~13 and ~32 RFU% activity reduction of MMP-9 and MMP-3, respectively, following 4-day incubation *in vitro*, whilst displaying an averaged mass loss in the range of 8-21 wt.%.

Dichroic and electrophoretic patterns of native type I collagen could still be observed following introduction of HA adducts, suggesting preserved triple helix architecture and chemical sequence in respective HA-methacrylated collagen conjugate. No hydrogel-induced toxic response was observed following 4-day culture of G292 cells, whilst lower compression modulus and gel content were measured in HA-bearing compared to methacrylated hydrogels, likely related to HA radical scavenging activity. The novel synthetic strategies described in this work provide a new insight into the systematic chemical manipulation of collagen materials aiming at the design of biomimetic, inflammation-responsive medical devices.

## 1. Introduction

Matrix metalloproteinases (MMPs) are a family of zinc and calcium dependent proteinases, which plays a key role in breaking down extracellular matrix (ECM) proteins. The levels of MMPs are precisely regulated and are responsible for key physiological events such as homeostasis, tissue remodelling, wound healing and angiogenesis [1-7]. Together with their beneficial role, numerous clinical studies have indicated that the overexpression of MMPs is associated with several inflammatory states, such as the ones found in chronic wounds, rotator cuff tears and cancer [8-13]. For example, elevated levels of MMPs are found in synovial fluid samples from the glenohumeral joints of patients with massive rotator cuff tears [14, 15]. Likewise, 5- to 10-fold increases in MMP levels were found in chronic leg ulcers compared to acute healing wounds; these elevated levels of MMPs explain the turnover in chronic tissue and failed wound closure [16-19]. Here, the imbalance between MMPs and tissue inhibitors of MMPs (TIMPs) promote constant degradation of neo-tissue and region-specific growth factors, so that the typical healing process is impaired. In light of the critical role MMPs play in maintaining the integrity of tissue ECM, growing research has therefore focused on the control of MMP expression aiming to manage inflammation and restore physiological homeostasis.

To date, 26 human MMPs have been found, which can be classified depending on the substrate specificity and molecular structure into at least four different classes, i.e. stromelysins (MMP-3, -10, -11, -7, -12), collagenases (MMP-1, -8, -13, -18), gelatinases (MMP-2 and -9) and membrane types MMPs (MMP-14, -15, -16, -17) [20]. MMPs, and especially collagenases, cleave all three α chains of native type I, II and

III collagens at a covalent bond between the glycine residue and the leucine or isoleucine residue, located approximately three-fourths of the way down the molecules from the N-terminus [21, 22]. Despite their classification, all MMPs present a catalytic zinc-binding domain, as well as a cysteine containing pro-peptide domain at the molecular level [23]. The cysteine residue is bound to the zinc atom when the enzyme is secreted in its latent conformation (pro-MMP). When the cysteine residue is dissociated from the zinc, the active site of the enzyme is exposed, triggering proteolytic activity. *In vivo*, proteolytic removal of the pro-domain for MMP activation is mediated by other proteinases, including MMPs, whilst MMP regulation can be further achieved by complexation with TIMPs, which block access to the active site [24].

Aiming to control inflammation-related MMP overexpression via systemic administration *in vivo*, the use of soluble factors has been widely investigated, whereby two main research approaches have been pursued:

(i) Synthesis of new TIMPs to restore physiological MMP/TIMP balance via systemic TIMP administration. However, despite several molecules having been synthesised and proposed for clinical use as TIMPs over the last 20 years, e.g. [4-(N-hydroxyamino)-2R-isobutyl-3S-(thiopen-2-ylthiomethyl)-succinyl]-L-phenylalanine-N-methylamide (Batimastat®) and N-[2,2-dimethyl-1-(methylcarbamoyl)propyl]-2-[hydroxy-(hydroxycarbamoyl)methyl]-4- methylpentanamide (Marimastat®), their clinical translation has been hampered by limited specificity and dose-limiting side effects (e.g. musculoskeletal syndrome) following systemic administration [25-27].

(ii) Application of chelating compounds aiming to revert MMP configuration to the native, deactivated state. Chelating agents, such as ethylenediaminetetraacetic acid (EDTA), have been proposed as zinc-chelating species to induce complexation with the exposed active site and to possess broad range MMP deactivation. Although this strategy has been successfully employed in some commercial wound dressings made from either hydrolysed collagen (Biostep®), oxidized regenerated cellulose and hydrolysed collagen (Promogran®), or a collagen-sodium alginate-carboxylmethylcellulose composite (ColActive®) [28, 29], challenges with respect to the controlled release of the chelating agents and long-term device functionality have to date been only partially addressed.

To avoid unwanted side and temporal effects and to achieve long-lasting MMP regulation *in situ*, the design of multifunctional polymer architectures has received growing attention aiming at soluble factor-free, device-based therapies [30, 31].

Synthetic polymers, i.e. polyacrylates [32] and respective hydroxamic acid-bearing derivatives [26], have been proposed as proof-of-concept anti-MMP substrates, due to the presence of ionically-charged and metal-chelating groups, respectively, along the polymeric backbone.

The concept of synthetic polymer-induced MMP complexation and proteolytic regulation has been transferred to inherently MMP-cleavable biopolymers, e.g. collagen and gelatin, to provide resulting medical devices with an additional biomimetic interface. Here, the amino acidic collagen phase of the medical device can act as alernative enzymatic substrate. According to this concept, the device-induced uptake of MMP-rich biological fluids triggers MMP-induced peptide bond cleavage. In this way, the MMP activity can be diverted to the collagen layer and away from the surrounding microenvironment, so that homeostasis can be restored. This concept has proved promising towards the development of device-based strategies for proteolytic activity regulation [33], although the rapid MMP-induced biodegradation may lead to device form-instability [34], non-controllable volumetric swelling, and poor mechanical properties *in vivo*, impacting on clinical performance.

These challenges have recently been addressed by the development of photo-curable crosslinking chemistries leading to tunable water-stable collagen networks. Photoinduced crosslinking proved key to realise collagen-based biomimetic systems with bespoke elasticity [35] and controlled triple helix organisation [36], whereby material biocompatibility and wound healing capability have been largely demonstrated [37-39]. UV-cured covalent networks made of functionalised collagen triple helices have also been presented with integrated MMP-chelating functions, whose swelling and compression properties can be controlled in a relatively independent fashion, via variation of e.g. type and extent of collagen functionalisation [40, 41]. On the other hand, current synthetic routes can only target a small content of easily-accessible functional groups, i.e. primary amines, along the collagen backbone, resulting in restricted customisation of material functionalities, e.g. with regards to MMP-modulation capability. In light of their chemical reactivity and low molar content, primary amino terminations of collagen are mainly employed for the generation of covalent crosslinks, so that the introduction of MMP-modulating chemical functions in resulting covalent networks is challenging to accomplish and control independently of the network crosslink density. Furthermore, currently-available crosslinking strategies of collagen still raises concerns with regards to the occurrence of irreversible, non-controllable crosslinking side reactions [43]. Consequently, flexible synthetic routes are still to be developed to fully realise multifunctional collagen systems with independently-customisable format, properties and functions.

In this study, we aimed to explore this challenge by building a hydroxamic acid-methacrylated collagen conjugate as a biomimetic building block for the synthesis of UV-cured hydrogel networks with MMP-modulating capability and independently-regulated crosslink density. Hydroxamic acid (HA) adducts were introduced via reaction with hydroxylamine, whereby HA well-known chelating activity was hypothesised to serve as means for MMP complexation and deactivation. Methacrylate functions were introduced via derivatisation of collagen amino groups to prompt the synthesis of a UV-cured crosslinked network in order to achieve enhanced enzymatic stability in the resulting materials. In addition to mediating the synthesis of the crosslinked network, the derivatisation of collagen with methacrylate functions was also exploited to protect the highly reactive, terminal amino groups of (hydroxy-)lysine residues to allow for the selective modification of carboxylic functions avoiding either irreversible and non-controllable side reactions or time-consuming de-/protection work-up.

Further to the quantification of the degree of collagen methacrylation via 2,4,6-trinitrobenzenesulfonic acid (TNBS), a new method is proposed to assess HA-induced consumption of carboxylic groups via indirect carboxylic acid amination in respective methacrylated and HA-conjugated collagen.

## 2. Materials and methods

### 2.1 Materials

Rat tails were provided by the School of Dentistry, University of Leeds, UK. Type I collagen was isolated in-house from the rat tail tendons with acidic treatment [44]. Methacrylic anhydride (MA), hydroxylamine hydrochloride (HA), triethylamine (TEA) and 2-mecaptoethanol (2-ME) were purchased from Sigma-Aldrich. *N*-(3-dimethylaminopropyl)-*N*'-ethylcarbodiimide hydrochloride (EDC), Ninhydrin and *N*-hydroxysuccinimide (NHS) were purchased from Alfa-Aesar. 2-Hydroxy-1-[4-(2-hydroxyethoxy) phenyl]-2-methylpropan-1-one (I2959) was purchased from Fluorochem Limited (Glossop, UK). All the other chemicals were purchased from Sigma-Aldrich.

## 2.2 Collagen methacrylation

Methacrylated collagen was prepared as previously reported [40]. Briefly, type I collagen from rat tail (CRT) was dissolved (0.25 wt.%) in 17.4 mM acetic acid solution and stirred at 4°C until complete dissolution was obtained. 0.1 M NaOH solution was added to achieve a pH of 7.5. MA and TEA were added with a molar ratio of either 25 or 35 with respect to the molar content of primary amino functions (~ 3 ×10$^{-4}$ mol·g$^{-1}$) available along the collagen backbone. After 24 hours the reacting mixture was precipitated in 10-volume excess of pure ethanol for 8 hours, recovered by centrifugation and air dried.

## 2.3 Covalent coupling of HA adducts on to methacrylated collagen

Methacrylated collagen (0.6 wt.%) was vigorously stirred in 0.01 M calcium-free phosphate buffer solution (PBS) until complete dissolution. EDC and NHS were added to the obtained solution to activate carboxylic acid terminations of collagen ([EDC]·[COOH]$^{-1}$= 6; [EDC]=[NHS]; [COOH]= 1 mmol·g$^{-1}$) [45, 46]. Following 60-min activation, an equimolar (to EDC) content of 2-ME was added to quench EDC activation. After 10-min reaction, either 2 or 4 molar excess (with respect to collagen carboxyl functions) of hydroxylamine chloride was added to induce HA covalent coupling. After 24-hour reaction, the mixture was precipitated in 10-volume excess of pure ethanol for 8 hours and recovered by centrifugation. This process was repeated twice and the collagen precipitate air dried.

## 2.4 Grafting of collagen carboxyl groups with ethylene diamine

Methacrylated and respective HA-conjugated collagens were aminated with ethylene diamine (EDA). This was carried out to indirectly quantify (via TNBS and Ninhydrin) the degree of HA-mediated collagen functionalisation by measuring HA-induced consumption of collagen carboxylic groups. Either methacrylated or HA-conjugated collagen (0.6 wt.%) was vigorously stirred in 0.01 M calcium-free PBS and 17.4 mM acetic acid, respectively, until complete dissolution. EDC and NHS ([EDC]·[COOH]$^{-1}$= 6 mmol·g$^{-1}$; [EDC]=[NHS]) were added and each solution stirred for one hour before adding an equimolar EDC content of 2-ME. After 10 minutes, 50 molar excess of EDA was added. This allowed for the grafting of EDA on to activated carboxyl groups of either methacrylated or HA-conjugated collagen, resulting in the introduction of, and the conversion of carboxyl functions to, terminal amino groups.

Following 24-hour reaction, reacting solutions were precipitated in 10-volume excess of pure ethanol for 8 hours and recovered by centrifugation. Retrieved products were re-dissolved in either PBS or 17.4 mM acetic acid, and precipitated again prior to air drying.

## 2.5 (2,4,6)-Trinitrobenzenesulfonic acid (TNBS) and Ninhydrin assays

The TNBS assay was used to directly measure derivatisation of amino groups and to indirectly measure derivatisation of carboxyl groups into either methacrylate, EDA or HA groups, and also respective collagen functionalisation. Briefly, 11 mg of dry samples were mixed with 1 mL of 4 wt.% $NaHCO_3$ (pH 8.5) and 1 mL of 0.5 wt.% TNBS solution at 40 °C under mild shaking. After 4-hour reaction, 3 mL of 6 M HCL solution was added and the overall reacting mixture incubated at 60°C for one hour to dissolve any sample residues. The solution was then cooled to room temperature, mixed with 5 mL distilled water, and extracted three times with 15 mL diethyl ether to remove non-reacted TNBS reagent. All samples were read against a blank (not containing samples) with an UV-Vis spectrophotometer (Model 6305, Jenway) at 346 nm, the content of free amino groups and degree of crosslinking (*F*) were calculated according to equations 1 and 2, respectively, as follows:

$$\frac{\text{mol(Lys)}}{\text{g(collagen)}} = \frac{2 \times \text{Abs(346nm)} \times 0.02}{1.46 \times 10^4 \times b \times x} \qquad \textbf{(Eq. 1)}$$

$$F = 100 - \frac{mol(Lys)_{funct.}}{mol(Lys)_{coll.}} \times 100 \qquad \textbf{(Eq. 2)}$$

where *Abs (346nm)* is the UV absorbance value recorded at 346 nm, *2* is the dilution factor, *0.02* is the volume of sample solution (in litres), *1.46×10⁴* is the molar absorption coefficient for 2,4,6-trinitrophenyl lysine (in $M^{-1} \cdot cm^{-1}$), *b* is the cell path length (1 cm), *x* is the dry sample weight, whilst *mol(Lys)$_{coll}$* and *mol(Lys)$_{funct.}$* represent the total molar content of free amino groups in native and functionalised collagen, respectively. *(Lys)* is hereby used to recognise that lysines make the highest contribution to the molar content of collagen free amino groups, although contributions from hydroxylysines and amino termini are also taken into account.

The Ninhydrin assay was further used to confirm the degree of derivatisation of the collagen amino groups previously-obtained via TNBS. 11 mg of dry sample were mixed with 4 mL distilled water and 1 mL of 8 wt.% Ninhydrin solution in acetone. The mixture was then incubated at 100 °C for 15 min. To terminate the reaction, the mixture was cooled on ice and 1 mL of 50 % (w/v) ethanol was added. The amount of free amino groups was determined by reading the absorbance at 570 nm against a blank sample (UV-Vis, Model 6305, Jenway). A standard calibration curve was prepared by carrying out above-mentioned assay with known mass of collagen. Three replicates were used for both TNBS and Ninhydrin measurements. Data are presented as mean±SD.

## 2.6 UV-induced hydrogel formation

Methacrylated collagen and respective HA-conjugated products were dissolved (1.2 wt.%) in 0.01 M calcium-free PBS containing 1 wt.% 2-Hydroxy-4'-(2-hydroxyethoxy)-2-methylpropiophenone (I2959) photoinitiator. The solution was centrifuged at 3000 rpm to remove any air bubbles, followed by casting (600 µL·well$^{-1}$) on to a 24-well plate (Corning Costar) and UV curing (Spectroline, 346 nm) for 30 mins at each side of the dish. Formed hydrogels were rinsed with distilled water, dehydrated via ascending series of ethanol solutions and air dried.

## 2.7 Quantification of gel content

The gel content was measured to investigate the overall portion of the covalent hydrogel network insoluble in 17.4 mM acetic acid solution [47]. Dry collagen networks ($m_d$: 0.01 – 0.02 g) were individually incubated in 2 mL of 17.4 mM acetic acid solution for 24 hours. Resulting samples were further air dried and weighed. The gel content ($G$) was calculated by the following equation:

$$G = \frac{m_1}{m_d} \times 100 \qquad \text{(Eq. 3)}$$

where $m_1$ is the dry mass after the incubation. Three replicates were used and the data are presented as mean±SD.

## 2.8 Compression tests

Air-dried UV-cured samples were equilibrated in PBS via overnight incubation at room temperature. Resulting PBS-equilibrated hydrogel discs (Ø: 18 mm; h: 5-6 mm) were compressed at room temperature with a compression rate of 3 mm·min$^{-1}$ (Instron ElectroPuls E3000). A 250 N load cell was operated up to complete sample compression. Stress–strain curves were recorded and the compression modulus quantified as the slope of the plot linear region at 25–30% strain. Three replicates were employed for each collagen network composition. Data are presented as mean±SD.

## 2.9 MMP activity and enzymatic degradation study

Full-length human pro-MMP-3 (PF063, Merck Millipore, UK) and pro-MMP-9 (PF038, Merck Millipore, UK) were activated in TCNB buffer containing 50 mM Tris, 10 mM $CaCl_2$, 150 mM NaCl and 0.05% Brij-23 (w/v), pH 7.5) in the presence of p-aminophenyl mercuric acetate (AMPA) with a final concentration of 1 mM at 37°C for 24 hours and 2 hours respectively, according to manufacturer's guidance. Dry collagen networks (n=4) were individually placed in a 24-well plate containing activated solutions of either MMP-3 or MMP-9; the plate was then incubated at 37°C on a benchtop orbital shaker for up to 4 days. The concentration of activated MMPs was 60 ng·mL$^{-1}$ and was selected according to the MMP concentration found in either torn rotator cuff or chronic wound fluid [8]. Sample-free solution controls of either activated or non-activated MMPs were used as control. A standard commercial assay (Abcam, ab112146) was used to detect the MMP activity in the supernatants; additionally, the MMP-incubated samples were collected, dehydrated with increasing distilled water/ethanol series (0, 25, 50, 80 and 100 vol.% ethanol) and air-dried. The mass of the retrieved dried samples was measured ($m_4$, n=4) and the relative mass loss calculated as the equation shown below:

$$\mu_{rel} = \frac{m_4}{m_d} \times 100 \qquad \text{(Eq. 4)}$$

## 2.10 Circular dichroism

Circular dichroism (CD) spectra of native and functionalised samples were acquired with a Chirascan CD spectrometer (Applied Photophysics Ltd) using 0.2 mg·mL$^{-1}$ solutions in 17.4 mM acetic acid or 0.01 M PBS solution. Sample solutions (n=2) were collected in quartz cells of 1.0 mm path length, whereby CD spectra were obtained with 2.0 nm band width and 20 nm·min$^{-1}$ scanning speed. A spectrum of the 17.4 mM acetic acid control solution was subtracted from each sample spectrum.

$$\theta_{mrw,\lambda} = \frac{MRW \times \theta_\lambda}{10 \times d \times c} \qquad \text{(Eq.5)}$$

where $\theta_\lambda$ is the observed molar ellipticity (degrees) at wavelength $\lambda$, $d$ is the path length and $c$ is the concentration (0.2 mg·mL$^{-1}$). *MRW* is the mean residue weight and equals to 91 g·mol$^{-1}$ for amino acids [48].

**2.11 Sodium dodecyl sulphate-polyacrylamide gel electrophoresis (SDS-PAGE)**

The samples were dissolved (1 wt.%) in SDS sample buffer containing 160 mM Tris-HCl (pH 6.8), 2% SDS, 26% glycerol, and 0.1% bromophenol blue. Sample solutions were heated for 60 seconds at 90 °C. 30 µL of each sample solution were added onto 4% stacking gel wells and separated on 15% resolving gels (200 V, 45 min). Protein bands were visualized after 60 min staining (0.1 wt.% Comassie Blue, 12.5 vol.-% trichloroacetic acid) and 60 min treatment in water under mild shaking. The molecular weight of resulting bands was approximately by measuring the relative mobility of the standard protein molecular weight markers.

**2.12 Cell culture**

Cells from the G292 cell line were cultured in Dulbecco's modified Eagle's medium (DMEM), supplemented with 10% fetal bovine serum (FBS), 1% glutamine, and 2.5 mg mL$^{-1}$ penicillin-streptomycin, in a humidified incubator at 37 °C and 5% $CO_2$. Cells were passaged every 3 days with 0.25% trypsin/0.02% EDTA. Freshly-synthesised, UV-cured hydrogel samples were incubated in a 70 vol.% ethanol solution under UV light. Retrieved samples were washed in PBS three times, prior to cell seeding. G292 cells (8×10$^3$ cells·ml$^{-1}$) were seeded on top of the hydrogel (following UV disinfection) and incubated at 37 °C for 4 days. After incubation, the hydrogels (n=6) were washed with PBS (x 3) and transferred to a new 24-well plate before adding the dying agent of Calcein AM and Ethidium homodimer-1; the plate was then incubated for 20 minutes away from light. Finally, live and dead stained hydrogels were placed on to a glass slide for fluorescence microscopy imaging (Leica DMI6000 B). Control cells were cultured on tissue culture plastic (negative) and killed by 30 min-incubation in 70 vol.% methanol (positive). Other than live /dead staining, cell viability was assessed using Alamar Blue assay (ThermoFisher Scientific, UK) according to manufacturer's guidance.

**2.13 Statistical analysis**

Statistical analysis was carried out using OriginPro 8.5.1. Significance of difference was determined by one-way ANOVA and post-hoc Tukey test. A *p* value of less than 0.05 was considered to be significant different. Data are presented as mean ± SD.

## 3. Results and discussion

In the following, the design of a HA-methacrylated collagen conjugate will be presented as a building block for a biomimetic medical device with integrated MMP-modulating capability. The medical device is realised at the molecular scale via the synthesis of a UV-cured hydrogel network of methacrylated collagen triple helices, in which MMP-modulating capability is achieved via introduction of HA adducts to carboxyl groups of methacrylated collagen.

Sample nomenclature used in this work is as follows: methacrylated and respective HA-conjugated collagen precursors are coded as 'MAXX' and 'HAYY', respectively, whereby 'MA' and 'HA' indicate the type of collagen adduct; whilst 'XX' and 'YY' describe the molar ratios of respective monomer in the functionalisation reaction. EDA-reacted samples were coded as either 'MAXX-EDA' or 'HAYY-EDA', respectively. Collagen networks are coded as either 'MAXX*' or 'HAYY*', whereby '*' identifies the chemically-crosslinked state, whilst the other labels have the same meaning as previously-mentioned.

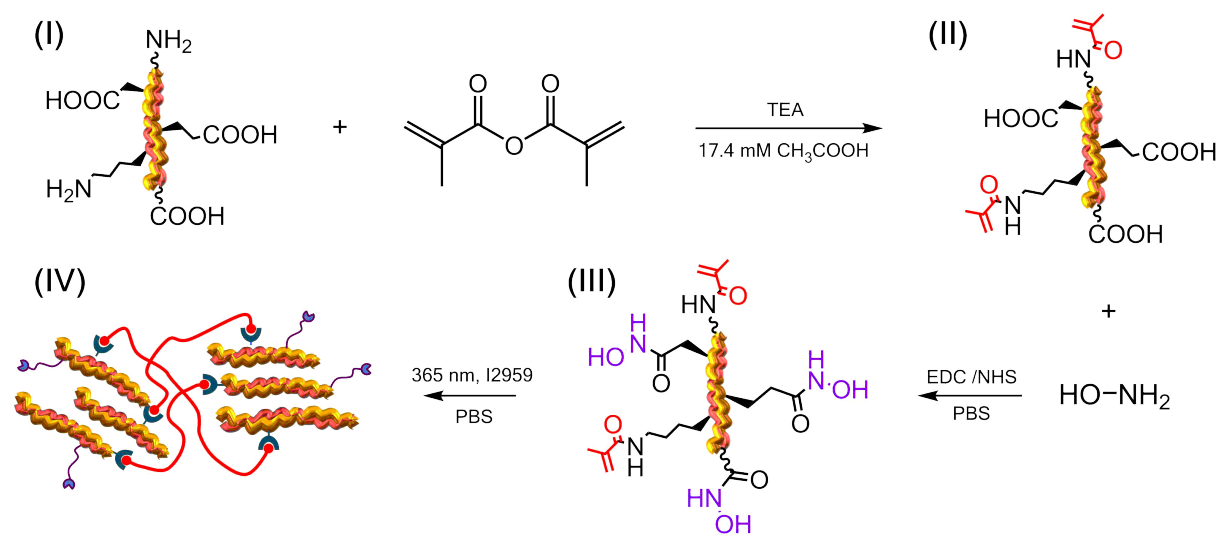

**Scheme 1.** Design of a soluble factor-free collagen system with integrated MMP-modulating capability. (I): Type I collagen is functionalised with MA with nearly-quantitative derivatisation of primary amino groups. (II): Retrieved product is reacted with hydroxylamine following carbodiimide-induced activation of collagen carboxylic groups, resulting in a photo-active collagen-HA conjugate with integrated MMP-chelating capability (III). (IV): UV irradiation in the presence of I2959 successfully leads to photo-induced crosslinking (—) of collagen triple helices ( ) bearing MMP-chelating HA adducts (—).

### 3.1 Synthesis of HA-methacrylated collagen conjugate

The TEA-catalysed reaction of MA with collagen proceeds via an amine-initiated nucleophilic addition/elimination mechanism and was therefore carried out prior to coupling with HA. This reaction with MA was exploited for two purposes (Scheme 1): (i) to introduce photoactive functions on to free amino terminations of collagen, e.g. (hydroxy-)lysines and amino termini, responsible for the subsequent formation of a hydrogel network [36, 40]; (ii) to protect highly-reactive amino terminations prior to selective derivatisation of carboxyl into MMP-chelating functional groups, so that intra-crosslinking side reactions could be minimised.

MA-mediated functionalisation of collagen was confirmed via TNBS and the content of MA adducts proved to be controlled depending on MA/Lys molar ration selected during the reaction (Table 1). TNBS assay of MA-reacted collagen samples revealed nearly-quantitative and reproducible consumption of primary amino groups, so that an averaged degree of functionalisation higher than 80 mol.% was observed ($F_{MA}$: 82 – 100 mol.%). Similar reactions have been described in the literature [40, 49, 50], although resulted in lower $F_{MA}$ values, suggesting that identified experimental conditions were crucial for the above findings. Although an $F_{MA}$ value of 100 mol.% was recorded in sample MA35, longer solubilisation time was required with respect to sample MA25 ($F_{MA}$: 94 mol.%); consequently, the latter product was selected for future investigations.

**Table 1.** Averaged degree of methacrylation ($F_{MA}$) determined via TNBS assay of collagen products (n=3) obtained with selected MA/Lys molar ration.

| Sample ID | $[MA][Lys]^{-1}$ | $F_{MA}$ /mol% |
|---|---|---|
| MA10 | 10 | 82 ± 2 |
| MA25 | 25 | 94 ± 1 |
| MA35 | 35 | 100 ± 0 |

To achieve the HA-methacrylated collagen conjugate, carbodiimide-induced activation of sample MA25 was pursued in PBS, due to the limited solubility of collagen triple helices in organic solvents. Introduction of EDC and NHS in collagen solutions is known to induce crosslinking reaction between collagen amino groups and NHS-activated collagen carboxylic groups, resulting in the formation of a gel [51]. In agreement with previously-discussed TNBS results, solution gelation was not observed in our case, providing evidence that minimal unwanted intra-crosslinking reaction occurred between residual amino groups and activated carboxyl groups of collagen. At the same time, solubilisation of retrieved, HA-reacted collagen product in an aqueous solution containing I2959 promptly led to the formation of a UV-cured hydrogel, as expected in light of the activation of covalently-coupled MA functions.

## 3.2 Quantification of HA-mediated functionalisation via an indirect amination strategy

Following reaction with HA and confirmation of UV-induced gelation, attention moved to the direct quantification of HA-mediated functionalisation in HA-methacrylated collagen samples. HA was covalently coupled to carboxylic acids of collagen, so that no direct colorimetric assay, i.e. TNBS or Ninhydrin, could be carried out with the resulting product. Kenawy et al. reported on the functionalisation of poly(N-acryloxysuccinimide) with HA, whereby resulting poly(N-hydroxyacrylamide) was successfully analysed via FTIR spectroscopy and $^1$H-NMR [52]. However, in the case of FTIR, the broad HA-related peaks at 3230 and 3480 cm$^{-1}$ (due to NH and OH bonds) overlap with the ones related to the stretching vibrations of collagen NH bonds. Likewise, HA-related $^1$H-NMR signals at $\delta$ 3.0 (singlet, OH) and $\delta$ 4.10 ppm (singlet, NH) are likely to overlap with the ones of the main collagen amino acids, e.g. proline. Consequently, assessment of HA content in HA-reacted collagen products can only be qualitative when using the above-mentioned methods.

In order to overcome this challenge, we aimed to *indirectly* quantify the molar content of HA by assessing the molar content of collagen carboxyl groups prior to and following reaction with HA, since the carboxyl groups in collagen were directly involved in the coupling reaction. To reach this goal, derivatisation of collagen carboxyl groups into TNBS- and Ninhydrin-detectable terminal amino groups was pursued in both methacrylated and HA-reacted collagen products via reaction with EDA (Scheme 2).

To avoid unwanted crosslinking reaction and promote selective grafting of EDA molecules, samples MA25, HA2 and HA4 were reacted with a 50-molar excess of EDA following carboxyl function activation with EDC/NHS. No gel formation was observed during the EDC-mediated activation of collagen and following reaction of EDC-activated collagen with EDA. These observations provided evidence that an EDA-grafted rather than crosslinked product was obtained, supporting the validity of the proposed approach in selectively derivitising carboxylic acids into terminal amines. Table 2 reports values related to the molar content of EDA- and HA-related adducts, as well as respective degrees of HA-mediated collagen functionalisation ($F_{HA}$).

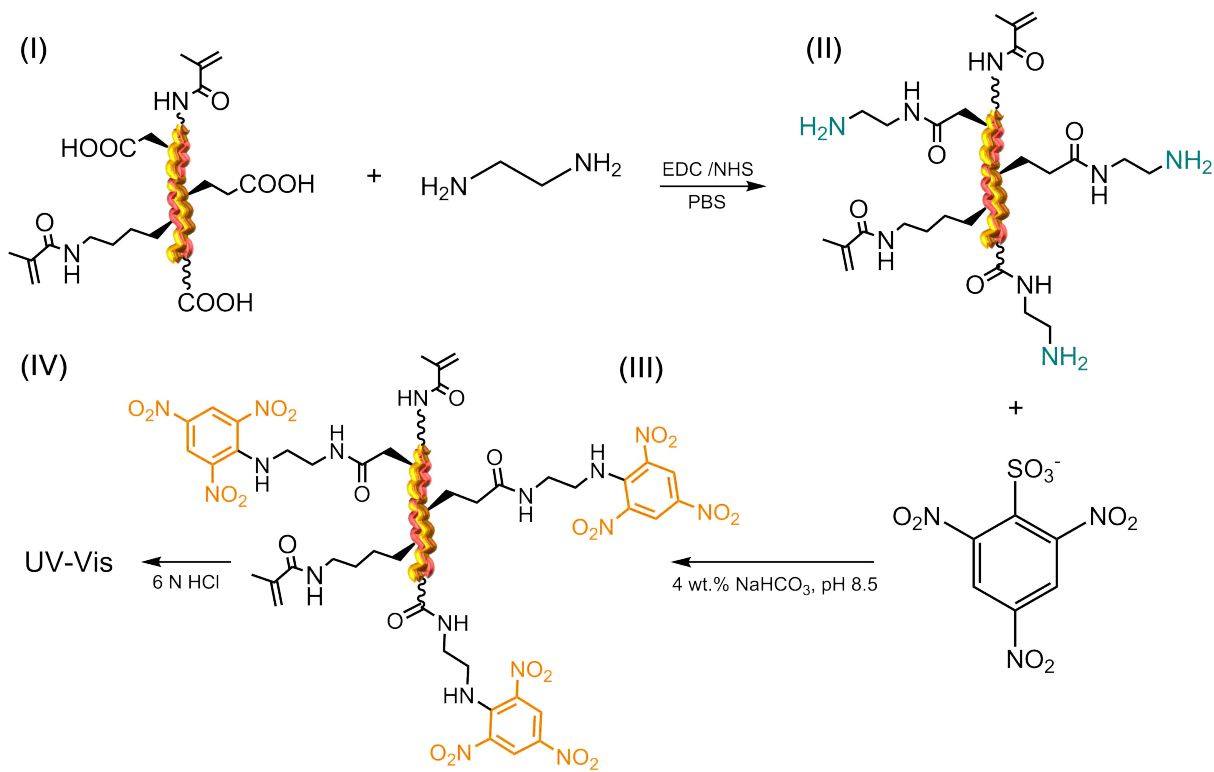

**Scheme 2.** Quantification of carboxyl terminations of collagen via an indirect amination strategy. Methacrylated collagen (I) was carbodiimide-activated and grafted with EDA, resulting in the direct conversion of carboxyl to primary amino functions (II). The aminated product was assessed with TNBS (III), allowing for the indirect quantification of pristine carboxyl functions via UV-Vis spectroscopy (IV).

EDA-induced grafting was confirmed by the fact that an increased content of free amino groups was recorded in sample MA25-EDA ([R-NH$_2$]: $1.80 \cdot 10^{-4}$ mol·g$^{-1}$) with respect to the one measured in sample MA25 ([Lys]: $0.19 \cdot 10^{-4}$ mmol·g$^{-1}$), so that an overall grafted EDA content of $1.61 \cdot 10^{-4}$ mol·g$^{-1}$ was successfully achieved. Most importantly, the molar content of amino groups was found to be lower in both HA-conjugated products, i.e. samples HA2-EDA ([EDA]: $0.14 \cdot 10^{-4}$ mol·g$^{-1}$) and HA4-EDA ([EDA]: $0.06 \cdot 10^{-4}$ mol·g$^{-1}$).

These results confirm the successful coupling of HA adducts on to collagen carboxyl functions. Considering an overall amount of carboxylic acids of $1 \times 10^{-3}$ mol·g$^{-1}$ in rat tail collagen [43, 44], above-mentioned results corresponded to an averaged degree of HA-mediated functionalisation of up to ~16 mol.% (Table 2). Together with TNBS, Ninhydrin assay was also carried out to further confirm the above findings. Also in this case, a decreased molar content of amino groups was recorded in HA-conjugated compared to methacrylated collagen samples. An averaged $F_{HA}$ value of up to ~14 mol.% was obtained, well in line with previous TNBS data.

**Table 2.** Degree of HA-mediated functionalisation ($F_{HA}$) determined in EDA-reacted, methacrylated, and respective HA-conjugated samples (n=3) via both TNBS and Ninhydrin assays. EDA terminal amino groups were used to indirectly quantify the consumption of carboxyl groups and consequent coupling of HA adducts. n.a.: not applicable.

| Sample ID | [EDA] /mol·g$^{-1}$ (×10$^{-4}$) | | [HA] /mol·g$^{-1}$ (×10$^{-4}$) | | $F_{HA}$ /mol.% (*) | |
|---|---|---|---|---|---|---|
| | TNBS | Ninhydrin | TNBS | Ninhydrin | TNBS | Ninhydrin |
| MA25-EDA | 1.61±0.08 | 1.48±0.05 | n.a. | n.a. | n.a. | n.a. |
| HA2-EDA | 0.14±0.04 | 0.17±0.03 | 1.47±0.04 | 1.21±0.08 | 14.7±0.4 | 12.1±0.7 |
| HA4-EDA | 0.06±0.02 | 0.09±0.04 | 1.55±0.06 | 1.37±0.09 | 15.5±0.6 | 13.7±0.9 |

(*) $F_{HA}$ was determined according to an overall content of collagen carboxyl groups of 1×10$^{-3}$ mol·g$^{-1}$ [53, 54].

Results obtained via collagen amination and colorimetric assays therefore provided supporting evidence that covalent coupling of both methacrylate and HA adducts could be selectively accomplished by independently targeting collagen amino and carboxyl groups, respectively. The nearly-quantitative lysine functionalisation with MA adducts proved key to selectively target free carboxyl functions of methacrylated collagen in a controlled fashion. The presented synthetic approach proved therefore proved reliable to enable the quantification of free carboxylic groups avoiding either the occurrence of well-reported, undesired intramolecular crosslinking reactions between collagen amino and carboxyl functions, or the employment of time-consuming de-/protection work-up [53, 55-57]. Although low variation in HA coupling was measured by both TNBS and Ninhydrin assays, a significant effect on the MMP-modulating capability of resulting UV-cured hydrogels was observed (section 3.4). It is expected that a wider range of functionalisation can be obtained by further adjusting the molar ratio of hydroxylamine with respect to carboxylic acid groups of methacrylated collagen.

### 3.3 Analysis of collagen conformation

SDS-PAGE was used to both elucidate the chemical composition of reacted products and to explore the triple helix organisation of methacrylated and HA-conjugated products. Retention of the triple helix architecture of collagen is key to enable resulting materials with chemotactic functionality, enhanced mechanical properties and decreased swellability [39]. Although we previously confirmed that collagen methacrylation had minimal effect on the collagen triple helix structure [40], the reaction of collagen-based polypetides with hydroxylamine has been reported to catalyse the chemical cleavage of peptide bonds between asparagine and glycine residues [58-60], which is likely to induce a denaturation of collagen triple helices. Despite the low (up to 9 mM) hydroxylamine concentrations employed in this study, it was important to confirm that no degradation product was formed in reacted species, aiming not to compromise chemical sequence and organisation of native collagen.

In-house extracted type I rat tail collagen was confirmed to display electrophoretic bands of monomeric α-chains (~100 kDa) and dimeric β-components (~200 kDa), as shown in Figure 1 (A). Each of these bands was observed in the electrophoretic patterns of samples MA25, HA2 and HA4, whilst no further band could be identified. SDS-PAGE data therefore supported the fact that no detectable collagen degradation occurred during the reaction with hydroxylamine. The minimal pattern variation in the electrophoretic reference bands associated with reacted, with respect to, native collagen, also suggested no alteration to the native triple helix organisation following collagen functionalisation.

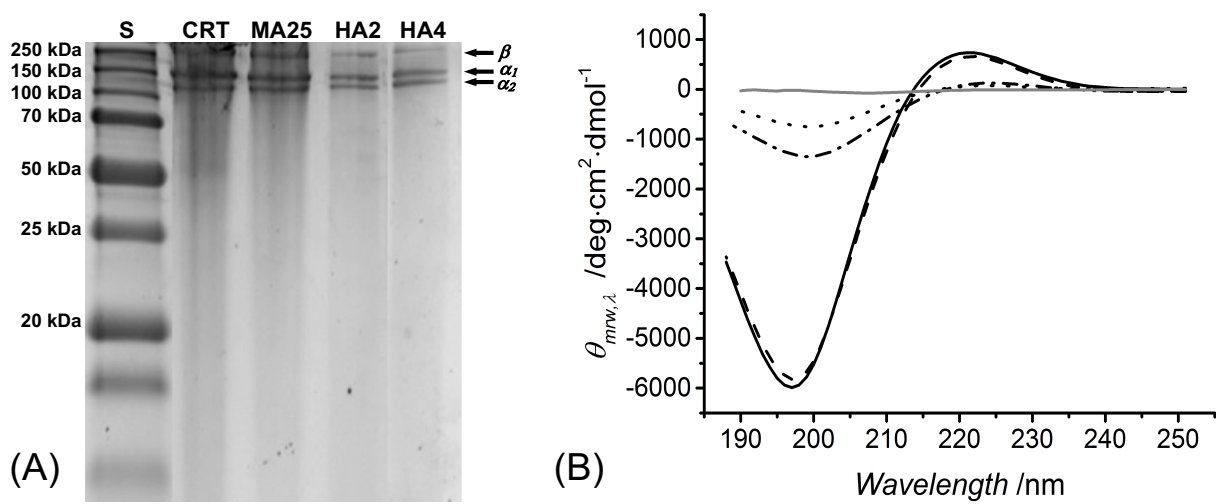

**Figure 1**. (A): SDS-PAGE analysis of standard (S), in-house isolated type I collagen from rat tail (CRT), as well as samples of MA25, HA2 and HA4. (B): Far-UV CD spectra (n=2) of samples of collagen (**black solid line**, *RPN* ~0.122), MA25 (**black dash line**, *RPN* ~0.113), HA2 (**black dash and dot**, *RPN* ~0.094), HA4 (**black dot**, *RPN* ~0.101) and gelatin (**grey**).

To further elucidate this point, CD spectroscopy was carried out. Collagen presents a unique CD spectrum with a negative peak at 197 nm and a positive peak at 221 nm [36, 61, 62]; these characteristic peaks could be identified in all collagen samples (Figure 1 B), confirming the presence of polyproline-II and triple helices, respectively [63]. The magnitude ratio of positive to negative peak RPN) in the CD spectra provides an indication of the content of collagen triple helices [40]. The RPN values measured in the CD spectra of samples MA25, HA2 and HA4 were 0.113, 0.094 and 0.101, respectively, which were only slightly lower than the RPN value measured in the CD spectra of native collagen (RPN: 0.122). By normalising the RPN value measured from functionalised samples with respect to the one of native collagen, a degree of triple helix preservation of at least 77 RPN% is observed. These results provide supporting evidence that collagen triple helices could still be preserved in HA-conjugated collagen samples, although at lower extent with respect to the case of methacrylated samples.

The preservation of collagen triple helices in HA-conjugated collagen is an interesting finding, given that additional HA, besides MA, adducts are covalently coupled to the collagen backbone. This finding may be explained by the fact that the overall HA molar content was limited up to only 16 mol.% of collagen carboxyl functions, suggesting that increasing the degree of HA-mediated functionalisation may negatively affect the triple helix organisation of native collagen.

### 3.4 Hydrogel impact on MMP activities and related enzymatic degradability

Following elucidation of the molecular architecture, protein organisation, and UV-cured network formation, hydrogel proteolytic degradability and MMP-regulation capability were studied *in vitro* with either MMP-3 or MMP-9. Both MMPs were selected in light of their documented overexpression in inflammation states, such as in the case of rotator cuff tear [5], chronic wounds [3, 9] and osteoarthritis [64]. Figure 2 (A) reports the MMP-3 activity of aqueous media following 4-day conditioning with samples MA25*, HA2* and HA4*, with respect to sample-free MMP-3 solutions.

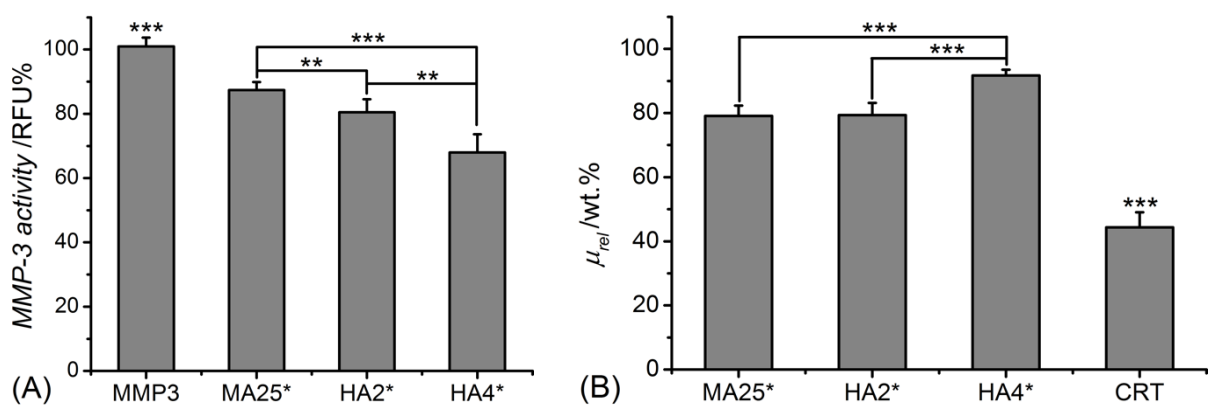

**Figure 2**. (A): MMP-3 activity measured *in vitro* following 4-day incubation with either no sample (MMP-3) or samples of MA25*, HA2* and HA4*. Results are normalised to the MMP-3 activity recorded in sample-free medium. (B): Relative mass of samples of CRT, MA25*, HA2* and HA4* following 4-day incubation in MMP-3-containing aqueous solution (60 ng·mL$^{-1}$ MMP-3). *$p<0.05$, **$p<0.01$ and ***$p<0.001$ are considered significant (n=4).

All samples proved to induce a significant reduction in MMP-3 activity ($p<0.001$), with samples HA2* (MMP-3: 81±4 RFU%) and HA4 (MMP-3: 70±6 RFU%) displaying significantly higher effect ($p<0.01$) with respect to methacrylated hydrogels (MMP-3: 87±2 RFU%). Together with analysis on supernatants, the relative mass of 4-day incubated samples was measured and compared to that of a control sample of native, non-crosslinked CRT (Figure 2 B). Hydrogel HA4* proved to show the highest relative mass ($\mu_{rel}$: 96±2 wt.%), which was significantly higher than the ones recorded in samples HA2* ($\mu_{rel}$: 79±4 wt.%) and MA25* ($\mu_{rel}$: 79±3 wt.%). As expected, the lowest relative mass was found in the control sample ($\mu_{rel}$: 44±4 wt.%), due to the absence of a crosslinked network at the molecular scale.

Similar trends were confirmed following sample incubation with MMP-9-containing aqueous media (Figure 3 A). MMP-9 activities were found to be reduced to either 88±5 or 87±4 RFU% following supernatant conditioning with either sample of HA2* or HA4*, with significant difference observed in comparison to supernatant treated with sample MA25* (MMP-9: 98±1 RFU%). As observed with MMP-3, sample CRT exhibits the lowest relative mass ($\mu_{rel}$: 57±7 wt.%), whilst both MMP-chelating hydrogels displayed more than 90 wt.% averaged relative mass (Figure 3 B). In comparison, sample MA25* displayed a lower relative mass ($\mu_{rel}$: 87±3 wt.%), which proved significantly different with respect to the ones of all crosslinked samples.

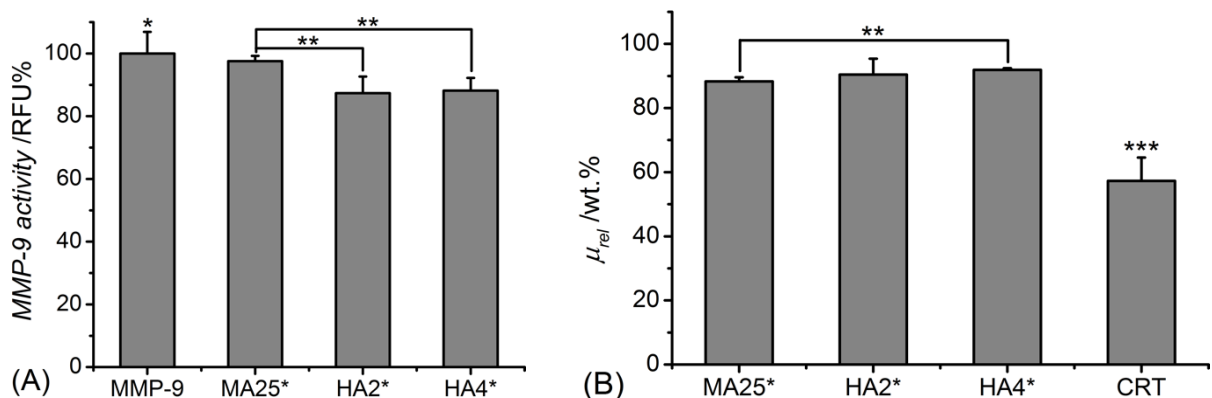

**Figure 3**. (A): MMP-9 activity measured *in vitro* following 4-day incubation with either no sample (MMP-9) or samples of MA25*, HA2* and HA4*. Results are normalised to the MMP-9 activity recorded in sample-free medium. (B): Relative mass of samples of CRT, MA25*, HA2* and HA4*, following 4-day incubation in MMP-9-containing aqueous solution (60 ng·mL$^{-1}$ MMP-9). *$p<0.05$, **$p<0.01$ and ***$p<0.001$ are considered significant (n=4).

MMP-3 and MMP-9 belong to two different classes of MMPs, namely stromelysins and gelatinases, respectively. Accordingly, MMP-3 presents broad substrate specificity and induces the activation of collagenases, e.g. MMP-1, a potent collagen-degrading enzyme. In contrast, MMP-9 is reported to preferentially degrade denatured collagen, i.e. gelatin, rather than collagen triple helices [65]. Our gravimetric data support the broader substrate specificity presented by MMP-3, whereby lower relative mass was recorded in both native collagen and hydrogel samples following incubation with MMP-3, compared to the case of the MMP-9–incubated samples. The broad specificity of MMP-3 compared to MMP-9 was also reflected by the higher hydrogel-induced activity reduction of the former compared to the latter enzyme. In light of MMP-9 specificity towards single rather than triple helices of collagen, above-mentioned observations provide further indirect evidence of triple helix retention in both methacrylated and HA-conjugated collagen hydrogels. The non-preferential specificity of MMP-9 towards collagen triple helices was also in agreement with the insignificantly different MMP-9 activity (Figure 3 A) and relative sample mass (Figure 3 B) recorded following 4-day incubation with either sample HA4* or sample HA2*, despite the increased HA content measured in the former compared to the latter system (Table 2). This observation correlated with previously-reported CD spectra (Figure 1 B), which indicated an increased triple helix-related RPN value in spectra of sample HA4 (RPN: 0.101), with respect to the case of sample HA2 (RPN: 0.094). Other than MMP-9, the fact that obtained hydrogel samples promoted increased deactivation of MMP-3 may be beneficial to control the excessive activation of tissue-detrimental collagenases [24]. Overall, above-mentioned proteolytic and gravimetric results with both MMP-3 and MMP-9 demonstrated the MMP-modulating capability induced by the functionalisation of the collagen backbone with MMP-chelating HA adducts. In contrast to endogenous MMP inhibitors, the HA-conjugated collagen hydrogels obtained in this study induce proteolytic modulation by binding to the active Zn site and a non-active Ca site inducing the MMP to adapt its native conformation.

Until now, there has been limited study on targeting the overexpression of MMPs due to the undesired side effects associated with the administration of soluble MMP inhibitors. Although HA-conjugated synthetic polymer systems have been presented in the past, no collagen variant has yet been developed, likely related to the challenges associated with the control and manipulation of the collagen architecture and the occurrence of crosslinking side reactions during collagen modification. Skarja *et al.* reported HA-containing microspheres prepared via derivatisation of poly(methyl methacrylate-co-methacrylic acid) and respective microsphere-induced reduction of the activity of MMP-2, -3, -8 and -13 [26]. Other studies investigated the effect of ovine-based collagen dressing on a broad spectrum of MMPs reduction but led to undesired clinical outcomes [66-69]. Layer by layer SiRNA coated nylon bandage was reported to yield rapid chronic wound closure in a diabetic mice model silencing 60% of MMP-9 activity in a two week study [70]; however, the biocompatibility of nylon may need to be considered further to enable the clinical applicability of a resulting device.

We have recently reported a 4VBC-functionalised collagen hydrogel with covalently-bound electron rich aromatic adducts, such that MMP-9 activity was reduced by 50% in 4 days *in vitro* [38]. With respect to that study, the hereby presented HA-conjugated formulation is likely to offer additional advantages, given the possibility to control the MMP-modulating capability independently of the crosslink density of the network.

### 3.5 Characterisation of physical properties

Once the MMP-modulating functionality of HA-bearing hydrogels was addressed, attention moved to the physical characterisation of the hydrogels. Gel content was quantified gravimetrically to gain information on the crosslink density and overall portion of non-extractable crosslinked material (Figure 4 A). Methacrylated hydrogels MA25* revealed the highest gel content of over 90 wt.% ($G$: 93±1 wt.%), which was higher than the one measured with samples HA2* ($G$: 87±1 wt.%) and HA4* ($G$: 85±5 wt.%).

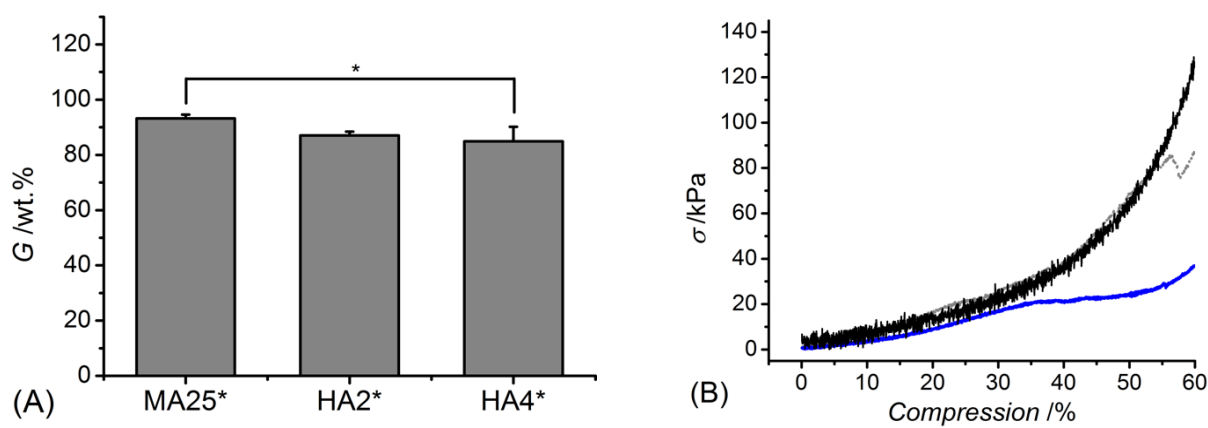

**Figure 4**. Gel content ($G$, A) and representative wet-state stress-compression curves (B) of samples MA25* (black), HA2* (grey) and HA4* (blue). *$p<0.05$ is considered significant (n=3).

Interestingly, despite the lower gel content in HA-based compared to methacrylated networks, samples HA* could still display decreased mass loss and induce increased reduction of MMP activity, further confirming the effect of covalently-coupled HA adducts on to the collagen backbone. In line with the gel content data, the compression modulus (Figure 4 B) proved to be significantly affected by the hydrogel formulation, so that sample MA25* showed significantly higher compressive modulus ($E_c$: 128±17 kPa) with respect to samples HA2* ($E_c$: 81±10 Pa) and HA4* ($E_c$: 72±16 kPa).

HA derivatives are known to display radical scavenging functionality [71], suggesting that the presence of HA adducts in the methacrylated collagen precursor impacts on the half-life of UV-generated radicals, crosslinking reaction kinetics and network crosslink density. This hypothesis is likely to explain the trends in gel content across the different formulations, since decreased gel content and compression modulus were observed in HA-bearing compared to methacrylated samples. The indirectly-observed radical scavenging functionality of HA-bearing hydrogels could also play a major role in controlling the excessive upregulation of reactive oxygen species found in e.g. chronic wounds, providing an additional capability in controlling inflammation.

### 3.6 *In vitro* cytotoxicity

The potential use of the HA-conjugated collagen hydrogels in biological system was evaluated by investigating the cellular response of osteosarcoma cells when cultured in direct contact with the material. Cell metabolism and viability were measured by Alamar blue assay at day 4 and 7 of cell culture (Figure 5), and exemplarily confirmed via live /dead staining on cells seeded on HA-conjugated collagen network HA4* following 4-day culture (Figure 6).

At day 4, the highest and lowest cell viability were recorded (via Alamar blue assay) in cells cultured onto either tissue culture plastic (TCP) or native collagen, whilst intermediate viability levels were seen in cells cultured with samples MA25*, HA2* and HA4* (Figure 5). No significant difference in cell viability was observed among the three UV-cured sample groups compared to the TCP group, whilst all of them were found to display increased cellular tolerance with respect to the sample of native collagen. At day 7, the highest cellular viability was measured via Alamar blue assay on cells cultured onto both TCP control and MA25* sample, with no statistical difference between them. Similar trends were also recorded in cells cultured onto native collagen control as well as onto samples of HA2* and HA4*, although respective cell viability proved to be significantly lower than the one recorded in TCP and MA25* groups.

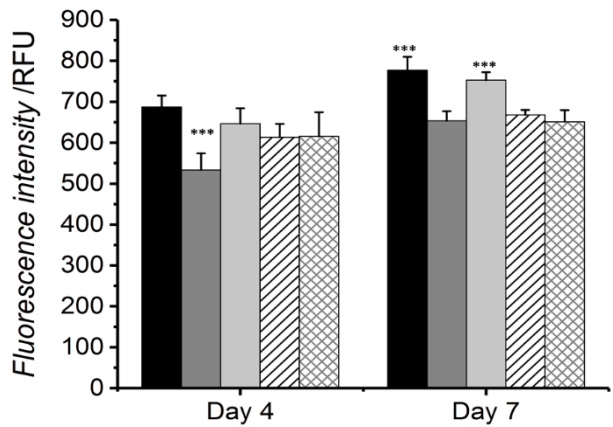

**Figure 5.** Alamar blue assay carried out with G292 osteosarcoma cells over 7 days. Cells were cultured on to either tissue culture plastic (black) and native collagen (grey) controls or samples of MA25* (light grey), HA2* (black, sparse patterned) and HA4* (grey, sparse patterned). *** indicates significantly-different means with respect to the other means at each specific time point ($p<0.001$, n=6).

Interestingly, sample MA25* seemed to induce a similar level of cellular proliferation and metabolism in comparison to the TCP control and in contrast to HA-conjugated collagen networks; this result may be due to the increased mechanical strength of hydrogel MA25* with respect to hydrogels HA2* and HA4*. Overall, these cell culture results successfully proved the high tolerability of all hydrogels tested in this study with G292 osteosarcoma cells.

In agreement with the results obtained via Alamar blue assay, no hydrogel-induced toxic response was confirmed via live/dead staining of G292 cells following 4-day culture onto sample HA4*. In contrast to the positive control confirming the presence of dead (red) cells (Figure 6 A), only vital (green) cells were indeed detected in samples HA4* (Figure 6 C-D), as in the case of the negative control (Figure 6 B).

Many reports suggested that soluble HA derivatives exhibit cytotoxicity against several cell lines such as human fibroblast, leukaemia and cancer cells [72-75]. This finding was not observed in this study as high cell viability was observed following 7-day culture onto HA-bearing collagen hydrogels with respect to native collagen controls. This observation further supports the use of HA-*coupled* rather than HA-*loaded* systems for the prolonged and localised regulation of upregulated MMPs.

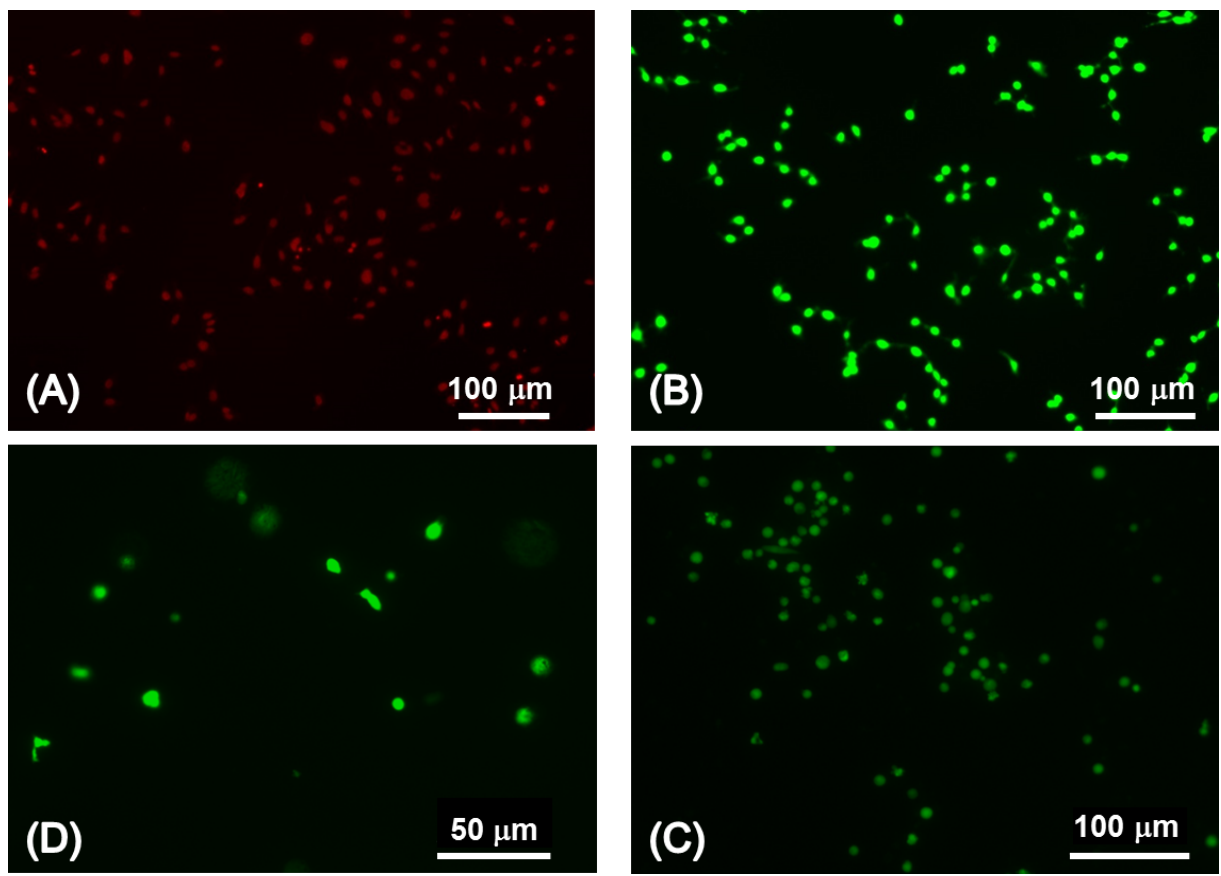

**Figure 6.** Typical live /dead staining of G292 osteosarcoma cells following 4-day culture. (A) and (B) show the positive and negative controls at day 4 of cell culture, respectively. (C) and (D) describe the live /dead staining of G292 cells following 4-day seeding on hydrogel HA4* at the magnification of ×10 and ×20, respectively.

## 4. Conclusions

The synthesis of a HA-methacrylated collagen conjugate was successfully demonstrated for the development of soluble factor-free hydrogel photonetworks with integrated MMP-regulating capability. Sequential covalent coupling of methacrylate *and* HA adducts to collagen was selectively accomplished by independently targeting amino and carboxyl groups, respectively. In this way, the hydrogel MMP-regulating capability could be controlled independently of the collagen network crosslink density. Amination of carboxyl groups via grafting reaction with EDA proved successful to indirectly quantify the degree of HA-mediated collagen functionalisation by both TNBS and Ninhydrin assays. Resulting HA-conjugated collagen samples maintained their unique triple helical structure, although at a lower extent with respect to methacrylated samples. The hydrogel-induced reduction in activities of both MMP-3 and MMP-9 demonstrated the MMP-modulating capability induced by the covalent functionalisation of the collagen backbone with MMP-chelating HA adducts, whereby the inherent MMP substrate specificity was observed to play a role. Increased enzymatic stability and decreased compressive modulus were measured in HA-conjugated compared to methacrylated hydrogel networks, in agreement with the presence of covalently-coupled HA adducts. Such materials were also shown to be cell friendly when cultured with G292 osteosarcoma cells. The versatility of this HA-methacrylated collagen conjugate has great potential in bypassing the well-known clinical problems associated with chelating agent-loaded medical products and provides a novel solution for the device-based modulation of MMP overexpression in a variety of diseases.

**Acknowledgements**

The authors gratefully acknowledge financial support provided by the EPSRC Centre for Innovative Manufacturing in Medical Devices (MeDe Innovation), as well as the Clothworkers' Centre for Textile Materials Innovation for Healthcare (CCTMIH). Jackie Hudson and Sarah Myers are gratefully acknowledged for technical assistance with fluorescence microscopy as well as cell culture facilities.

**References**

1. Tan, R.J. and Y. Liu, *Matrix metalloproteinases in kidney homeostasis and diseases.* American Journal of Physiology - Renal Physiology, 2012. **302**(11): p. F1351-F1361.
2. Gaffney, J., et al., *Multilevel regulation of matrix metalloproteinases in tissue homeostasis indicates their molecular specificity in vivo.* Matrix Biology, 2015. **44**: p. 191-199.
3. Caley, M.P., V.L.C. Martins, and E.A. O'Toole, *Metalloproteinases and Wound Healing.* Advances in Wound Care, 2015. **4**(4): p. 225-234.
4. Armstrong, D.G. and E.B. Jude, *The role of matrix metalloproteinases in wound healing.* J Am Podiatr Med Assoc, 2002. **92**(1): p. 12-8.
5. Jacob, J., et al., *Matrix metalloproteinase levels as a marker for rotator cuff tears.* Orthopedics, 2012. **35**(4): p. e474-8.
6. De Giorgi, S., M. Saracino, and A. Castagna, *Degenerative disease in rotator cuff tears: what are the biochemical and histological changes?* Joints, 2014. **2**(1): p. 26-28.

7. Gialeli, C., A.D. Theocharis, and N.K. Karamanos, *Roles of matrix metalloproteinases in cancer progression and their pharmacological targeting.* Febs j, 2011. **278**(1): p. 16-27.
8. Jacob, J., et al., *Matrix Metalloproteinase Levels as a Marker for Rotator Cuff Tears.* ORTHOPEDICS, 2012. **35**: p. 474-478.
9. Ravanti, L. and V.M. Kahari, *Matrix metalloproteinases in wound repair (review).* Int J Mol Med, 2000. **6**(4): p. 391-407.
10. Osawa, T., T. Shinozaki, and K. Takagishi, *Multivariate analysis of biochemical markers in synovial fluid from the shoulder joint for diagnosis of rotator cuff tears.* Rheumatology International, 2005. **25**(6): p. 436-441.
11. Whittaker, M., et al., *Design and Therapeutic Application of Matrix Metalloproteinase Inhibitors.* Chemical Reviews, 1999. **99**(9): p. 2735-2776.
12. Posnett, J. and P.J. Franks, *The burden of chronic wounds in the UK.* Nurs Times, 2008. **104**(3): p. 44-5.
13. Yoshihara, Y., et al., *Biochemical markers in the synovial fluid of glenohumeral joints from patients with rotator cuff tear.* Journal of Orthopaedic Research, 2001. **19**(4): p. 573-579.
14. Dodson, C.C., et al., *The long-term outcome of recurrent defects after rotator cuff repair.* Am J Sports Med, 2010. **38**(1): p. 35-9.
15. Nagra, N.S., et al., *Mechanical properties of all-suture anchors for rotator cuff repair.* Bone & Joint Research, 2017. **6**(2): p. 82-89.
16. Wysocki, A.B., L. Staiano-Coico, and F. Grinnell, *Wound Fluid from Chronic Leg Ulcers Contains Elevated Levels of Metalloproteinases MMP-2 and MMP-9.* Journal of Investigative Dermatology, 1993. **101**(1): p. 64-68.
17. Yager, D.R., et al., *Wound Fluids from Human Pressure Ulcers Contain Elevated Matrix Metalloproteinase Levels and Activity Compared to Surgical Wound Fluids.* Journal of Investigative Dermatology, 1996. **107**(5): p. 743-748.
18. Muller, M., et al., *Matrix metalloproteinases and diabetic foot ulcers: the ratio of MMP-1 to TIMP-1 is a predictor of wound healing.* Diabetic Medicine, 2008. **25**(4): p. 419-426.
19. Rayment, E.A., Z. Upton, and G.K. Shooter, *Increased matrix metalloproteinase-9 (MMP-9) activity observed in chronic wound fluid is related to the clinical severity of the ulcer.* British Journal of Dermatology, 2008. **158**(5): p. 951-961.
20. Gibson, D., et al., *MMPs made easy.* Wounds International, 2009. **1**(1): p. 1-6.
21. Song, F., et al., *Matrix metalloproteinase dependent and independent collagen degradation*. Vol. 11. 2006. 3100-20.
22. Salsas-Escat, R., P.S. Nerenberg, and C.M. Stultz, *Cleavage Site Specificity and Conformational Selection in Type I Collagen Degradation.* Biochemistry, 2010. **49**(19): p. 4147-4158.
23. Van Wart, H.E. and H. Birkedal-Hansen, *The cysteine switch: a principle of regulation of metalloproteinase activity with potential applicability to the entire matrix metalloproteinase gene family.* Proceedings of the National Academy of Sciences of the United States of America, 1990. **87**(14): p. 5578-5582.
24. Gooyit, M., et al., *A Chemical Biological Strategy to Facilitate Diabetic Wound Healing.* ACS Chemical Biology, 2014. **9**(1): p. 105-110.
25. Purcell, B.P., et al., *Injectable and bioresponsive hydrogels for on-demand matrix metalloproteinase inhibition.* Nat Mater, 2014. **13**(6): p. 653-661.
26. Skarja, G.A., et al., *The effect of a hydroxamic acid-containing polymer on active matrix metalloproteinases.* Biomaterials, 2009. **30**(10): p. 1890-1897.

27. Lin, Y.-A., et al., *Rational Design of MMP Degradable Peptide-Based Supramolecular Filaments.* Biomacromolecules, 2014. **15**(4): p. 1419-1427.
28. Brett, D., *A Review of Collagen and Collagen-based Wound Dressings.* Wounds, 2008. **20**(12): p. 347-56.
29. Fisher, J.F. and S. Mobashery, *Recent advances in MMP inhibitor design.* Cancer and Metastasis Reviews, 2006. **25**(1): p. 115-136.
30. Sefton, M., M. May, and G. Skarja, *Hydroxyamate-containing materials for the inhibition of matrix metalloproteinases.* 2004, Google Patents.
31. Renò, F., V. Traina, and M. Cannas, *Adsorption of matrix metalloproteinases onto biomedical polymers: a new aspect in biological acceptance.* Journal of Biomaterials Science, Polymer Edition, 2008. **19**(1): p. 19-29.
32. Eming, S., et al., *The inhibition of matrix metalloproteinase activity in chronic wounds by a polyacrylate superabsorber.* Biomaterials, 2008. **29**(19): p. 2932-2940.
33. Hart, J., et al., *The role of oxidised regenerated cellulose/collagen in wound repair: effects in vitro on fibroblast biology and in vivo in a model of compromised healing.* The International Journal of Biochemistry & Cell Biology, 2002. **34**(12): p. 1557-1570.
34. Qiao, X., et al., *Compositional and in Vitro Evaluation of Nonwoven Type I Collagen/Poly-dl-lactic Acid Scaffolds for Bone Regeneration.* Journal of Functional Biomaterials, 2015. **6**(3): p. 667.
35. Pierce, B.F., et al., *Photocrosslinked Co‐Networks from Glycidylmethacrylated Gelatin and Poly(ethylene glycol) Methacrylates.* Macromolecular Bioscience, 2012. **12**(4): p. 484-493.
36. Tronci, G., S.J. Russell, and D.J. Wood, *Photo-active collagen systems with controlled triple helix architecture.* Journal of Materials Chemistry B, 2013. **1**(30): p. 3705-3715.
37. Ouasti, S., et al., *Network connectivity, mechanical properties and cell adhesion for hyaluronic acid/PEG hydrogels.* Biomaterials, 2011. **32**(27): p. 6456-6470.
38. Tronci, G., et al., *Protease-sensitive atelocollagen hydrogels promote healing in a diabetic wound model.* Journal of Materials Chemistry B, 2016. **4**(45): p. 7249-7258.
39. Holmes, R., et al., *Thiol-Ene Photo-Click Collagen-PEG Hydrogels: Impact of Water-Soluble Photoinitiators on Cell Viability, Gelation Kinetics and Rheological Properties.* Polymers, 2017. **9**(6): p. 226.
40. Tronci, G., et al., *Multi-scale mechanical characterization of highly swollen photo-activated collagen hydrogels.* Journal of The Royal Society Interface, 2015. **12**(102).
41. Tronci, G., et al., *Influence of 4-vinylbenzylation on the rheological and swelling properties of photo-activated collagen hydrogels.* MRS Advances, 2015. **1**(8): p. 533-538.
42. Mohammadi, H., et al., *Inelastic behaviour of collagen networks in cell–matrix interactions and mechanosensation.* Journal of The Royal Society Interface, 2015. **12**(102).
43. Van Nieuwenhove, I., et al., *Protein functionalization revised: N-tert-butoxycarbonylation as an elegant tool to circumvent protein crosslinking.* Macromol Rapid Commun, 2014. **35**(15): p. 1351-5.
44. Rajan, N., et al., *Preparation of ready-to-use, storable and reconstituted type I collagen from rat tail tendon for tissue engineering applications.* Nat. Protocols, 2007. **1**(6): p. 2753-2758.

45. Brown, E.M., H.M. Farrell, Jr., and R.J. Wildermuth, *Influence of neutral salts on the hydrothermal stability of acid-soluble collagen.* J Protein Chem, 2000. **19**(2): p. 85-92.
46. Mayne, J. and J.J. Robinson, *Comparative analysis of the structure and thermal stability of sea urchin peristome and rat tail tendon collagen.* Journal of Cellular Biochemistry, 2002. **84**(3): p. 567-574.
47. Jin, R., et al., *Injectable chitosan-based hydrogels for cartilage tissue engineering.* Biomaterials, 2009. **30**(13): p. 2544-2551.
48. Nishio, T. and R. Hayashi, *Regeneration of a Collagen-like Circular Dichroism Spectrum from Industrial Gelatin.* Agricultural and Biological Chemistry, 1985. **49**(6): p. 1675-1682.
49. Brinkman, W.T., et al., *Photo-Cross-Linking of Type I Collagen Gels in the Presence of Smooth Muscle Cells: Mechanical Properties, Cell Viability, and Function.* Biomacromolecules, 2003. **4**(4): p. 890-895.
50. Gaudet, I.D. and D.I. Shreiber, *Characterization of Methacrylated Type-I Collagen as a Dynamic, Photoactive Hydrogel.* Biointerphases, 2012. **7**(1): p. 25.
51. Tronci, G., et al., *Triple-helical collagen hydrogels via covalent aromatic functionalization with 1,3-Phenylenediacetic acid.* Journal of materials chemistry. B, Materials for biology and medicine, 2013. **1**(40): p. 5478-5488.
52. Kenawy, E.-R., et al., *A New Degradable Hydroxamate Linkage for pH-Controlled Drug Delivery.* Biomacromolecules, 2007. **8**(1): p. 196-201.
53. Everaerts, F., et al., *Quantification of carboxyl groups in carbodiimide cross-linked collagen sponges.* Journal of Biomedical Materials Research Part A, 2007. **83A**(4): p. 1176-1183.
54. Brown, E.M., H.M. Farrell, and R.J. Wildermuth, *Influence of Neutral Salts on the Hydrothermal Stability of Acid-Soluble Collagen.* Journal of Protein Chemistry, 2000. **19**(2): p. 85-92.
55. Van Nieuwenhove, I., et al., *Protein Functionalization Revised: N-tert-butoxycarbonylation as an Elegant Tool to Circumvent Protein Crosslinking.* Macromolecular Rapid Communications, 2014. **35**(15): p. 1351-1355.
56. Staros, J.V., R.W. Wright, and D.M. Swingle, *Enhancement by N-hydroxysulfosuccinimide of water-soluble carbodiimide-mediated coupling reactions.* Analytical Biochemistry, 1986. **156**(1): p. 220-222.
57. Sell, S.A., et al., *Cross-linking methods of electrospun fibrinogen scaffolds for tissue engineering applications.* Biomed Mater, 2008. **3**(4): p. 045001.
58. Hu, J., et al., *Chemical cleavage of fusion proteins for high-level production of transmembrane peptides and protein domains containing conserved methionines.* Biochimica et Biophysica Acta (BBA) - Biomembranes, 2008. **1778**(4): p. 1060-1066.
59. Crimmins, D.L., S.M. Mische, and N.D. Denslow, *Chemical Cleavage of Proteins in Solution*, in *Current Protocols in Protein Science*. 2001, John Wiley & Sons, Inc.
60. Saris, C.J.M., et al., *Hydroxylamine cleavage of proteins in polyacrylamide gels.* Analytical Biochemistry, 1983. **132**(1): p. 54-67.
61. Holmgren, S.K., et al., *Code for collagen's stability deciphered.* Nature, 1998. **392**(6677): p. 666-667.
62. Lopes, J.L.S., et al., *Distinct circular dichroism spectroscopic signatures of polyproline II and unordered secondary structures: Applications in secondary *structure analyses.* Protein Science : A Publication of the Protein Society, 2014. **23**(12): p. 1765-1772.
63. Drzewiecki, K.E., et al., *Circular Dichroism Spectroscopy of Collagen Fibrillogenesis: A New Use for an Old Technique.* Biophysical Journal, 2016. **111**(11): p. 2377-2386.
64. Yang, C.-C., et al., *Matrix Metalloproteases and Tissue Inhibitors of Metalloproteinases in Medial Plica and Pannus-like Tissue Contribute to Knee Osteoarthritis Progression.* PLoS ONE, 2013. **8**(11): p. e79662.
65. Ågren, M.S., et al., *Tumor necrosis factor-α-accelerated degradation of type I collagen in human skin is associated with elevated matrix metalloproteinase (MMP)-1 and MMP-3 ex vivo.* European Journal of Cell Biology, 2015. **94**(1): p. 12-21.
66. Negron, L., S. Lun, and B.C.H. May, *Ovine forestomach matrix biomaterial is a broad spectrum inhibitor of matrix metalloproteinases and neutrophil elastase.* International Wound Journal, 2014. **11**(4): p. 392-397.
67. Bohn, G., et al., *Ovine-Based Collagen Matrix Dressing: Next-Generation Collagen Dressing for Wound Care.* Advances in Wound Care, 2016. **5**(1): p. 1-10.
68. Chattopadhyay, S. and R.T. Raines, *Review collagen-based biomaterials for wound healing.* Biopolymers, 2014. **101**(8): p. 821-833.
69. Brett, D., *A Review of Collagen and Collagen-based Wound Dressings*. Vol. 20. 2015. 347-56.
70. Castleberry, S.A., et al., *Self-Assembled Wound Dressings Silence MMP-9 and Improve Diabetic Wound Healing In Vivo.* Advanced Materials, 2015: p. n/a-n/a.
71. Končić, M.Z., et al., *Antiradical, Chelating and Antioxidant Activities of Hydroxamic Acids and Hydroxyureas.* Molecules, 2011. **16**(8): p. 6232.
72. Przychodzen, W., et al., *Cytotoxic and Antioxidant Activities of Benzohydroxamic Acid Analogues*. Vol. 34. 2013.
73. Librizzi, M., et al., *The Histone Deacetylase Inhibitor JAHA Down-Regulates pERK and Global DNA Methylation in MDA-MB231 Breast Cancer Cells.* Materials, 2015. **8**(10): p. 7041-7047.
74. Nagaoka, Y., et al., *Synthesis and cancer antiproliferative activity of new histone deacetylase inhibitors: hydrophilic hydroxamates and 2-aminobenzamide-containing derivatives.* European Journal of Medicinal Chemistry, 2006. **41**(6): p. 697-708.
75. Angibaud, P., et al., *Discovery of pyrimidyl-5-hydroxamic acids as new potent histone deacetylase inhibitors.* European Journal of Medicinal Chemistry, 2005. **40**(6): p. 597-606.